\documentclass[]{siamart1116}

\usepackage{amssymb,amsfonts,amsmath}

\usepackage[mathscr]{eucal}

\usepackage{stmaryrd}

\usepackage{amsbsy}

\usepackage{bm}

\usepackage{braket}

\usepackage{xspace}

\usepackage{multirow}

\usepackage{tabularx}
\newcolumntype{Y}{>{\raggedright\arraybackslash}X}

\usepackage{xparse}

\usepackage{xfrac}

\usepackage{environ}

\usepackage{tikz}
\usetikzlibrary{3d}
\usetikzlibrary{patterns}
\usetikzlibrary{calc}
\usetikzlibrary{arrows}

\usepackage{pgfplots}

\usepackage[font=footnotesize,justification=Centering,singlelinecheck=false]{subfig}

\usepackage{cleveref}

\usepackage{algpseudocode}

\Crefname{ALC@unique}{Line}{Lines}

\NewDocumentCommand \qtext {m} {\quad\text{#1}\quad}

\NewDocumentCommand \Real {} {\mathbb{R}}

\DeclareDocumentCommand \i {} {i}

\NewDocumentCommand \X {} {\boldsymbol{\mathscr{X}}}

\NewDocumentCommand \Xn { O{n} } {\mathbf{X}_{(#1)}}

\DeclareDocumentCommand \xi { s } 
{
  \IfBooleanTF{#1}
  {x(i_1,i_2,\dots,i_d)}
  {x_{\i}}
}

\DeclareDocumentCommand \Xi { s }{X_{\i}}

\NewDocumentCommand \M {} {\boldsymbol{\mathscr{M}}}

\NewDocumentCommand \Mk { O{k} } {\mathbf{M}_{(#1)}}

\DeclareDocumentCommand \mi { s } 
{
  \IfBooleanTF{#1}
  {m(i_1,i_2\dots,i_d)}
  {m_{\i}}
}

\NewDocumentCommand \An { G{n} t' } {%
  \IfBooleanTF{#2}%
  {\bigl(\mathbf{A}^{(#1)}\bigr)^{\intercal}}%
  {\mathbf{A}^{(#1)}}  
}

\NewDocumentCommand \Bn { G{n} } {%
  \mathbf{B}^{(#1)}
}

\NewDocumentCommand \akikj { O{k} G{i} G{j} } {
  a_{#1}(#2_{#1},#3)
}

\NewDocumentCommand \akij { O{k} G{i_k} G{j} } {
  a_{#1}(#2,#3)
}

\NewDocumentCommand \KT { s } {
  \llbracket 
  \IfBooleanTF{#1}{{\bm\lambda};}{}
  \An{1}, \An{2}, \dots,  \An{d} \rrbracket
}

\NewDocumentCommand \W {} {\boldsymbol{\mathscr{W}}}

\DeclareDocumentCommand \wi { s } 
{
  \IfBooleanTF{#1}
  {w(i_1,i_2\dots,i_d)}
  {w_{\i}}
}

\NewDocumentCommand \G {} {\boldsymbol{\mathscr{G}}}

\DeclareDocumentCommand \gi { s } 
{
  \IfBooleanTF{#1}
  {g(i_1,i_2\dots,i_d)}
  {g_{\i}}
}

\NewDocumentCommand \Gk { O{k} } {\mathbf{G}_{(#1)}}

\NewDocumentCommand \Zk { G{k} } {\mathbf{Z}_{#1}}

\NewDocumentCommand \Z { } {\mathbf{Z}}

\NewDocumentCommand \V { } {\mathbf{V}}

\NewDocumentCommand \I {} {\mathcal{I}}

\NewDocumentCommand \FPD { s m m } {
  \IfBooleanTF{#1}
  {\partial #2 / \partial #3}
  {\frac{\partial #2}{\partial #3}}
}

\NewDocumentCommand \genten {} {\textrm{GenTen}\xspace}

\usepackage{soul}

\NewDocumentCommand \VerA {} {MTTKRP-A\xspace}
\NewDocumentCommand \VerB {} {MTTKRP-B\xspace}
\NewDocumentCommand \VerC {} {MTTKRP-C\xspace}

\usepackage{graphicx}
\usepackage{epstopdf}
\usepackage{amsopn}
\ifpdf
  \DeclareGraphicsExtensions{.eps,.pdf,.png,.jpg}
\else
  \DeclareGraphicsExtensions{.eps}
\fi
\usepackage{listings}
\definecolor{lightlightgray}{gray}{0.97}
\definecolor{lightlightgreen}{rgb}{0.8,1,0.8}
\definecolor{darkblue}{rgb}{0,0,0.8}
\definecolor{darkgreen}{rgb}{0,0.5,0}
\def\inline{\lstinline[basicstyle=\ttfamily,keywordstyle={}]}

\usepackage{tabulary}
\usepackage{booktabs}
\usepackage{subfig}
\usepackage{placeins}
\usepackage{afterpage}

\numberwithin{theorem}{section}

\newcommand{\TheTitle}{Software for Sparse Tensor Decomposition on Emerging Computing Architectures} 

\newcommand{\TheAuthors}{E. Phipps and T. Kolda}

\headers{Sparse Tensor Decomposition on Emerging Architectures}{\TheAuthors}

\title{{\TheTitle}\thanks{
    {This work was supported by the Laboratory Directed Research and Development Program at Sandia National Laboratories.  Sandia National Laboratories is a multimission laboratory managed and operated by National Technology and Engineering Solutions of Sandia LLC, a wholly owned subsidiary of Honeywell International Inc., for the U.S. Department of Energy's National Nuclear Security Administration under contract DE-NA0003525.  This paper describes objective technical results and analysis. Any subjective views or opinions that might be expressed in the paper do not necessarily represent the views of the U.S. Department of Energy or the United States Government.}}}

\author{
  Eric Phipps\thanks{Sandia National Laboratories, Albuquerque, NM
    (\email{etphipp@sandia.gov}).}
  \and
  Tamara G. Kolda\thanks{Sandia National Laboratories, Livermore, CA
    (\email{tgkolda@sandia.gov}).}
}

\ifpdf
\hypersetup{
  pdftitle={\TheTitle},
  pdfauthor={\TheAuthors}
}
\fi

\begin{document}

\expandafter\def\expandafter\UrlBreaks\expandafter{\UrlBreaks%
  \do\a\do\b\do\c\do\d\do\e\do\f\do\g\do\h\do\i\do\j%
  \do\k\do\l\do\m\do\n\do\o\do\p\do\q\do\r\do\s\do\t%
  \do\u\do\v\do\w\do\x\do\y\do\z\do\A\do\B\do\C\do\D%
  \do\E\do\F\do\G\do\H\do\I\do\J\do\K\do\L\do\M\do\N%
  \do\O\do\P\do\Q\do\R\do\S\do\T\do\U\do\V\do\W\do\X%
  \do\Y\do\Z}

\maketitle

\begin{abstract}
In this paper, we develop  software for decomposing sparse tensors that is {\em portable to and performant on} a variety of multicore, manycore, and GPU computing architectures. The result is a single code whose performance matches optimized architecture-specific implementations. The key to a portable approach is to determine multiple levels of parallelism that can be mapped in different ways to different architectures, and we explain how to do this for the matricized tensor times Khatri-Rao product (MTTKRP) which is the key kernel in canonical polyadic tensor decomposition. Our implementation leverages the Kokkos framework, which enables a single code to achieve high performance across multiple architectures that differ in how they approach fine-grained parallelism. We also introduce a new construct for portable thread-local arrays,
  which we call compile-time polymorphic arrays. Not only are the specifics of our approaches and implementation interesting for tuning tensor computations, but they also provide a roadmap for developing other portable high-performance codes. As a last step in optimizing performance, we modify the MTTKRP algorithm itself to do a permuted traversal of tensor nonzeros to reduce atomic-write contention. We test the performance of our implementation on 16- and 68-core Intel CPUs and the K80 and P100 NVIDIA GPUs, showing that we are competitive with state-of-the-art architecture-specific codes while having the advantage of being able to run on a variety of architectures.
\end{abstract}


%

\section{Introduction}
\label{sec:intro}
%

Tensors, or multidimensional arrays, are a powerful means of representing relationships in multiway data \cite{KoBa09}.
We focus on computing the canonical polyadic or CANDECOMP/PARAFAC (CP) decomposition~\cite{Ha70,CaCh70} for sparse tensors.
CP decompositions have numerous applications in data science, including analysis of online social networks \cite{GuPaPa17}, anomaly detection \cite{FaGa16a}, compression of neural nets \cite{JaSeAn15a,NoPoOsVe15,CoShSh16}, and health data analytics \cite{WaChGhDe15}, among others.
The CP decomposition is a low-rank decomposition and approximates a given tensor by a sum of rank-one tensors.
In this work, our focus is on the main computational kernel in computing the CP decomposition:
the matricized~tensor~times~Khatri-Rao~product (MTTKRP).
We describe the mathematical background on tensors and MTTKRP in \cref{sec:cp}.

Our goal is to develop CP decomposition software that is {\em portable to and performant on} a variety of multicore and manycore computing architectures, such as multicore
CPUs, the manycore Intel Xeon Phi, and NVIDIA GPUs.  This means we desire software that runs on a variety of computing architectures, contains few architecture-specific optimizations,  ports to new architectures with only small code modifications, and yet achieves performance on par with customized implementations that have been optimized for each architecture.
Relying on a single code implementation simplifies software development and maintenance while providing some degree of ``future-proofing'' as new architectures and programming models are developed.

The key to developing portable code is to identify multiple levels of parallelism that can be mapped in different ways to different architectures.
For MTTKRP with a sparse tensor, we observe parallelism across the nonzeros and also across the columns of the factor matrices.
To realize the portable parallelism in practice, we leverage the Kokkos package~\cite{Kokkos:2012:SciProg,Kokkos:2014:JPDC}, which provides threading abstractions that allow a single software implementation
to be portable to a diversity of shared memory parallel programming models (such as OpenMP, pThreads, and CUDA) and hardware.
For example, fine-grained parallelism on CPU and Xeon Phi architectures is expressed through vector arithmetic operations that are usually incorporated through automatic vectorization of low-level loops by the compiler.  Conversely, fine-grained parallelism on GPUs is expressed through the explicit programming of groups of cooperating threads.
Kokkos 
provides abstractions that unify these different approaches, making a \emph{single} software implementation leveraging fine-grained parallelism feasible.  The multiple levels of parallelism and use of Kokkos is described in \cref{sec:cp-als-kokkos}.

One issue with Kokkos is that it does not provide a good way to allocate thread-private arrays.
To address this, we introduce
a new portable and performant data abstraction based on compile-time
polymorphic arrays in \cref{sec:comp-time-polym},
and we contrast this with Kokkos scratch pad arrays. 
Our method enables better vector parallelism at the lowest loop level, especially on GPUs, 
and, we contend,
is simpler to use than what was already available in Kokkos.
Using polymorphic arrays for fine-grained parallelism does not depend on Kokkos and could also be used in other contexts.

To avoid race conditions in thread-parallel writes in MTTKRP, we rely on atomic write instructions since they are generally scalable for high concurrency architectures.  Nevertheless, these atomic instructions can substantially reduce performance for many tensors, so a modification to the MTTKRP algorithm is described in \cref{sec:perm-mttkrp} that reorders operations to reduce contention among threads writing to the same memory location,
with the tradeoff being storage of permutation arrays that doubles the size of the tensor stored in memory.

Our portable and performant software implementation of the CP alternating least squares (CP-ALS) method for sparse tensors is available in the open source \genten library.\footnote{\url{https://gitlab.com/tensors/genten}}
\Cref{sec:results} provides performance results for  this code with the different versions of MTTKRP discussed in the paper.
We test on multicore CPUs, NVIDIA GPUs, and the Knights Landing (KNL) version of the Intel Xeon Phi.
We also compare our implementation to a state-of-the-art open-source code called SPLATT~\cite{SmRaSiKa15}. 
These results show that \genten using the modified MTTKRP algorithm achieves state-of-the-art performance on a variety of platforms with a single codebase.

To summarize, this work provides three primary contributions.  First and foremost, this work describes how to achieve performance portability for the MTTKRP kernel by identifying coarse and fine-grained parallelism and leveraging the Kokkos package to exploit it.
Second, we introduce a novel construct that uses
thread-local compile-time polymorphic arrays for leveraging
fine-grained parallelism in  MTTKRP.
Third, we present a new permuted variant of MTTKRP that
dramatically reduces costs associated with atomic writes.
Taken together, our publicly available \genten software library incorporating these contributions represents the first-ever implementation of CP-ALS in a portable manner,  achieving high performance on CPUs, Intel Xeon Phi, and NVIDIA GPUs with a single implementation.

\section{Related work}
\label{sec:related-work}

SPLATT~\cite{SmRaSiKa15} stores each mode of a sparse tensor as a list of slices, where each slice is stored in a compressed format similar to the compressed sparse row/compressed row storage (CSR/CRS) format of sparse matrices.  The MTTKRP algorithm is parallelized over the rows of the result matrix using a task parallelism scheme implemented through OpenMP.  A cache blocking scheme is also introduced to improve MTTKRP performance.  Because of the mode-dependent storage format, this approach requires a different representation for each mode of the tensor which substantially increases memory costs.  {More recently, SPLATT incorporated a compressed sparse fiber (CSF) approach~\cite{Smith:2015ff} that stores a tensor as a family of trees, which in principle avoids duplication of the tensor for each mode.  However because a mode must be chosen to form the roots of the trees, the resulting MTTKRP algorithm can have substantially different performance depending on which mode is being traversed.  So in practice multiple CSF representations are computed and stored in memory to reduce total CP-ALS time.}  The MTTKRP algorithm incorporates thread parallelism and employs a tiling mechanism to avoid the use of locks or atomic instructions in order to handle thread race conditions.  The performance of the CSF approach in SPLATT has also been recently explored on the KNL version of the Intel Xeon Phi architecture~\cite{Smith:2017em}, where several variants of the MTTKRP algorithm were considered, including a thread privatization approach to handling race conditions.  {Conversely, our work considers the coordinate-based format for sparse tensor storage exclusively, which does not require any duplication of the tensor, and the use of atomic instructions for handling race conditions.  Since the use of atomics on some architectures (such as KNL) can be quite slow, we also propose a modified MTTKRP algorithm that substantially reduces atomic contention, with a little less than twice the memory storage.  We also consider portability to GPUs in our work.}

Building on the CSF data structure, Li at al.\@ proposed an adaptive tensor memoization algorithm that reduces the number of redundant floating-point operations that occur during the sequence of MTTKRP calculations required by CP-ALS, with the trade-off of increased memory usage~\cite{Li:2017fx} due to storing semi-sparse intermediate tensors.  An adaptive model tuning framework called AdaTM was also developed that chooses an optimized memoization algorithm based on the sparse input tensor.  Fine-grained parallelism across multiple modes of the intermediate tensors was proposed but only studied within the context of thread parallelism on multicore CPU architectures.  {Portability to KNL and GPU architectures is not considered.}

{DFacTo~\cite{ChVi14} stores a sparse tensor as a set of sparse matrices representing unfoldings of the tensor for each mode, and reorganizes MTTKRP for each mode as a sequence of sparse matrix-vector products (SpMV).  As such DFacTo can rely on optimized implementations of SpMV, and therefore focuses on distributed memory parallelism.  DFacTo doesn't consider portability explicitly, but reliance on a SpMV essentially provides it.  However DFacTo's MTTKRP algorithm has only been developed for third-order tensors and requires storing a separate unfolding for each tensor mode.  DFacTo stores these in a compressed CSR format which makes the storage reasonably memory efficient, but requires packing all but one of the coordinates for each nonzero into a single ordinal, which may not fit if the tensor has long mode lengths.  Furthermore, use of CSR-based SpMV for MTTKRP can be inefficient due to the usual row-based parallelization strategy if there is a wide spread in the number of nonzeros per tensor slice (resulting in imbalance in the number of nonzero columns in the unfolded tensor), or if the tensor has short mode lengths (resulting in insufficient parallelism).}

The Cyclops Tensor Framework~\cite{solomonik2014massively} provides a general framework for implementing dense and sparse tensor operations upon which CP decompositions can be built.  For sparse tensors, the library relies on matricization and a corresponding sparse matrix primitive, {and therefore suffers the same limitations as DFacTo described above.}  The library is geared towards distributed memory parallelism.

More recently, Liu et al.\@ proposed an extension of coordinate-based tensor storage format called F-COO (flagged-coordinate)~\cite{2017arXiv170509905L} which is extensible to many types of tensor operations, and demonstrated MTTKRP and CP decomposition results on GPUs.  {Portability to non-GPU architectures is not considered.  F-COO modifies the traditional coordinate-based format (COO) by adding bit arrays that determine when a mode index changes and are used to implement a thread-parallel MTTKRP algorithm based on segmented reduction, thereby avoiding atomic instructions.  The authors claim F-COO requires less memory storage than COO (since one index is replaced by bit arrays), but since MTTKRP operations are required for all modes in CP-ALS, storage of the bit flag arrays for all modes increases total memory storage by a small factor.  The authors also claim that an MTTKRP algorithm based on atomics is too slow on the GPU, but as we will show in \cref{sec:results}, the algorithm performs very well on newer GPU models with fast atomic hardware.  The F-COO approach is conceptually similar to the permutation-based approach we propose in \cref{sec:perm-mttkrp} which also reduces the costs associated with atomics.  Instead of using bit arrays to implement segmented reduction, we instead change the order of traversal of nonzeros using permutation arrays to make segmented reduction unnecessary.  Our approach also has the advantage of improving the locality of writes when updating the resulting factor matrix in MTTKRP, which while not important on GPUs, is quite important on cache-based architectures such as CPUs and the Xeon Phi.  However the F-COO approach does require somewhat less storage than the permutation approach we propose.}

 Finally, Li et al.\@ proposed a variant of the coordinate format called HiCOO~\cite{Li:2018sc} that decomposes a sparse tensor into small sparse blocks, reducing the memory required to store tensor nonzeros (and hence memory bandwidth to read them).  {A thread-parallel MTTKRP algorithm is developed for small thread-count architectures such as CPUs that groups the small blocks into larger blocks called superblocks with the collection of superblocks distributed across threads.  Race conditions are handled either by parallelizing between superblocks across the mode that is being written to when the length of that mode is large, or by using a thread privatization strategy when the mode length is small.  Portability to GPUs is not considered, and it is unclear how the algorithm would be parallelized on GPUs with very large numbers of threads.  Likely the best approach would be to forego the two-level blocking strategy and instead parallelize directly between the small tensor blocks using atomics for handling race conditions (at least for recent GPUs with fast atomics).  In this case, many of the portability ideas introduced in our work could be applied to this approach.}

%
%
%
%

\section{CP tensor decomposition and MTTKRP}
\label{sec:cp}
%

Except for specific relevant concepts discussed in detail later, we assume in this section a basic familiarity with tensors and refer the reader to
Kolda and Bader~\cite{KoBa09} for further details.
Let $\X \in \Real^{I_1 \times \cdots \times I_d}$ be a given $d$-way tensor.
We refer to each way or dimension as a \emph{mode}.
For a given rank $R$, the goal is to find a low-rank model tensor $\M$ that is a good approximation to $\X$, i.e., 
\begin{equation}\label{eq:CP}
  \min_{\M} \;\; \| \X-\M \| \quad \mbox{s.t.} \quad
  \M = \sum_{j=1}^R \lambda_j \, \mathbf{a}^{(1)}_j \circ \mathbf{a}^{(2)}_j \circ \dots \circ \mathbf{a}^{(d)}_j,
\end{equation}
where $\lambda_j$ is a scalar weight, $\mathbf{a}^{(n)}_j$ is a column vector of size $I_n$ that is assumed to be normalized to unit-norm in some norm,
and $\circ$ represents the tensor outer product.
For notational convenience, we assemble all the column vectors in mode $n$ into a \emph{factor matrix} of size $I_n \times R$:
\begin{displaymath}
  \An =
  \begin{bmatrix}
    \mathbf{a}^{(n)}_1 &  \cdots & \mathbf{a}^{(n)}_R
  \end{bmatrix}.
\end{displaymath}
Hence, the goal is to find the \emph{weight vector} $\bm{\lambda} =
\begin{bmatrix}
  \lambda_1 & \cdots & \lambda_R
\end{bmatrix}^{\intercal}
$ and the $d$ factor matrices $\{ \An{1}, \dots, \An{d} \}$ that define the
low-rank model tensor $\M$.
Following~\cite{BaKo07}, we refer to $\M$ as a Kruskal tensor, or K-tensor for short.

In this work, we compute the CP decomposition using the alternating
least squares~(CP-ALS) method~\cite{Ha70,CaCh70}.  Details are omitted
here but can be found in, e.g., the survey by Kolda and Bader
\cite{KoBa09}.  Our main interest is in the matricized tensor times
Khatri-Rao product (MTTKRP) calculation for a sparse tensor, so we
focus on this kernel for sparse tensors for the remainder of this section.

We say a tensor is \emph{sparse} if the majority of its elements are zero. 
We can store such a tensor efficiently by storing only its
nonzeros and their indices \cite{BaKo06}.
Here, we denote the $i$th nonzero and its subscripts as
$x_i$ and  $(\ell_{i1}, \ell_{i2},\dots,\ell_{id})$ respectively.
If there are $P$ nonzeros, then we store $\X$ with a $P$-vector of real values
and a $P \times d$ vector of coordinates.
As discussed in \cref{sec:related-work}, a variety of sparse tensor formats have been proposed in the literature \cite{BaKo07,SmRaSiKa15,ChVi14,2017arXiv170509905L,Li:2018sc}. 
Storing the tensor $\X$ as list nonzero indices and values~\cite{BaKo07} is called \emph{coordinate format} (COO), and this is what we use here since it does not favor any particular tensor mode.

For a sparse tensor in COO format, the mode-$n$ MTTKRP computes a matrix $\V$ of size $I_n \times R$ that is defined elementwise as \cite{BaKo07}
\begin{equation}
\label{eq:mttkrp}
  v(k,j) = \lambda_j \sum_{{i=1}\atop{\ell_{in} = k}}^P x_i \prod_{{m=1}\atop{m \neq n}}^d a^{(m)} (\ell_{im},j)
  \qtext{for}
  k = 1,\dots,I_n \text{ and } j = 1,\dots,R.
\end{equation}
We assume $d$ is small, ranging from 3 to 5 in the examples we show.
In order of magnitude, $P$ ranges from $10^6$--$10^8$, $I_n$ usually
ranges from $10^3$--$10^6$ but can be as small as 2, and $R$ usually
ranges from 10--100. The MTTKRP is the primary bottleneck of CP-ALS and the
primary focus of our parallelization efforts.  
%

%
%
%
%
%
%
%
%
%
%
%
%
%
%
%
%
%
%
%
%
%
%
%
%
%
%
%
%
%
%
%
%

%
%
%
%
%
%
%
%
%
%
%

%
%
%
%
%
%
%
%
%
%
%
%

%
%
%
%
%
%
%
%
%
%
%
%
%
%
%
%
%

%
%
%
%
%
%
%
%
%
%
%
%
%
%
%
%
%
%
%
%
%
%
%
%

%
%
%
%
%
%
%
%
%
%
%
%
%
%
%
%
%
%
%
%
%
%

%
%
%
%
%
%
%
%
%
%
%
%

%
%
%
%
%
%
%
%
%
%

%
%
%
%

\section{Parallelism in MTTKRP}
\label{sec:cp-als-kokkos}
%

In this section we describe the multi-level parallelism present in MTTKRP and a portable implementation.
Our goal is to be able to efficiently exploit fine-grained parallelism, by which we refer to the Single Instruction Multiple Data (SIMD) parallelism provided by vector instructions on  multicore CPUs and the manycore Intel Xeon Phi, as well as the Single Instruction Multiple Thread (SIMT) parallelism provided by the fine-grained threads within a warp on CUDA (NVIDIA) GPUs. 
Ideally, we can use a single code that runs on any architecture, which is referred to as \emph{portable}.
Our portable implementation uses
Kokkos~\cite{Kokkos:2012:SciProg,Kokkos:2014:JPDC}, a programming model and C++ library that enables applications and domain libraries to implement thread scalable algorithms that are efficient and portable across modern architectures.
To discuss the levels of parallelism in a way that is agnostic to specific architectures, we use terminology mirroring the nested parallelism concepts introduced in the OpenMP 4.0 specification~\cite{openmp_4_0_spec} consisting of a league of teams, where each team is comprised of a collection of threads, and each thread may execute vector instructions in parallel.  The league is virtual (i.e., not tied to any hardware resource) and corresponds to the highest level parallel iteration space, while a team within the league corresponds to a collection of one or more hyperthreads on a CPU/Phi architecture or a thread block on a GPU architecture, and vector parallelism corresponds to CPU/Phi vector instructions or threads within a GPU warp. See \cref{tab:parallelism} for a summary.

\begin{table}
  \centering\footnotesize
  \caption{Levels of Parallelism}
  \label{tab:parallelism}
  \begin{tabularx}{\linewidth}{|l|>{\raggedright\arraybackslash}p{1.75in}|>{\raggedright\arraybackslash}X|}
    \hline
    \bf Level & \bf CPU Interpretation & \bf GPU Interpretation \\ \hline
    League & Collection of teams & Grid (group of thread blocks), cannot synchronize \\  \hline
    Team & Collection of hyperthreads on one or more cores & Thread block (usually 128 to 256 threads), shares fast memory and can synchronize \\ \hline
    Vector & Vector instructions on single thread & Threads within GPU warp (i.e., 32 threads), execute in SIMD fashion \\ \hline
  \end{tabularx}
\end{table}

\subsection{Multi-level parallelism for MTTKRP}
\label{sec:mttkrp}

Our goal is to exploit as much available parallelism in the MTTKRP calculation as possible.  We use a multilevel parallelism approach, incorporating parallelism over tensor nonzeros, i.e., the $i$ index in \cref{eq:mttkrp}, and factor matrix columns, i.e., the $j$ index in \cref{eq:mttkrp}.  
For the latter we use vector-level parallelism, since it provides the best performance when iterating over contiguous regions in memory to enable packed/coalesced memory accesses,\footnote{On caching architectures such as CPUs and the Intel Xeon Phi, packed accesses refers to a given thread accessing consecutive memory locations sequentially and is a prerequisite for transforming a loop to use vector instructions.  Conversely, coalesced accesses on a GPU refers to consecutive threads accessing consecutive memory locations, which is required to achieve full memory bandwidth.} which is possible with the combination of a rowwise layout of the factor matrices and a row-oriented MTTKRP algorithm that processes (portions of) factor matrix rows simultaneously.
A pseudo-code description of the algorithm is shown in \cref{fig:mttkrp_alg}.
The variables should generally be the same as in \cref{eq:mttkrp}, but the  
less intuitive mappings between the source code variables and the math are given in \cref{tab:math_to_code}.
Several parameters are architecture dependent, and the values for those under different situations are shown in \cref{tab:arch_params}. 

\begin{figure}[ht]
\begin{lstlisting}
// Compute MTTKRP with sparse tensor X, using K-Tensor M, in mode n. 
// The result is stored in the factor matrix V.
mttkrp(Sptensor X, Ktensor M, unsigned n, FacMatrix V) {

  // Problem parameters
  P = X.nnz()              // Number of nonzeros
  d = M.ndims()            // Number of dimensions
  R = M.ncomponents()      // Number of factor matrix components  
  
  // Architecture-specific values
  vector_size = ...             // Vector size (see (*\cref{tab:arch_params}*))
  FBS = ...                     // Factor matrix column block size (see (*\cref{tab:arch_params}*))
  team_size = ...               // Team size (see (*\cref{tab:arch_params}*))
  NZPTM = 128                   // Nonzeros per team member
  NZPT = NZPTM * team_size      // Nonzeros per team
  league_size = (P+NZPT-1)/NZPT // Number of tensor nonzero blocks (league size)

  // Top level of parallelism...
  parallel_league_for(league_rank=0; league_rank<league_size; ++league_rank) {
    tmp = ScatchPad(team_size, FBS) // shared within the team
    
    // Loop over factor matrix column blocks
    for (jb=0; jb<R; jb+=FBS) {
      nj = (jb+FBS < R) ? FBS : R-jb  // number of columns to process

      // Second level of parallelism...
      parallel_team_for(team_rank=0; team_rank<team_size; ++team_rank) {
        i_offset = league_rank*NZPT + team_rank*NZPTM  // starting nonzero index
        ni = (i_offset+NZPTM < P) ? NZPTM : P-i_offset // number of nonzeros

        // Loop over tensor nonzeros in block
        for (i=i_offset; i<i_offset+ni; ++i) {
          x_val = X.value(i)     // value for nonzero i
          k = X.subscript(i,n)   // mode-n index for nonzero i
      
          // Initialize to x-value times lambda-weight
          parallel_vector_for(j=0; j<nj; ++j) 
            tmp(team_rank,j) = x_val * M.weights(jb+j)
          
          // Multiply by corresponding factor matrix entries
          for (m=0; m<d; ++m) 
            if (m != n) 
              parallel_vector_for(j=0; j<nj; ++j) 
                tmp(team_rank,j) *= M[m].entry(X.subscript(i,m),jb+j)
    
          // Multiple teams may be contributing to the same entries of the result
          parallel_vector_for(j=0; j<nj; ++j) 
            atomic_add(V.entry(k,jb+j), tmp(team_rank,j))

        } // i
      } // team_rank
    } // jb
  } // league_rank
} // function
\end{lstlisting}
  \caption{\textbf{\VerA .} Pseudo-code description of efficient parallel MTTKRP calculation in a C++-like syntax.
    Several of the algorithmic parameter choices are architecture dependent; see \cref{tab:arch_params}.
    It has three levels of parallelism at the league, team, and vector levels.}
\label{fig:mttkrp_alg}
\end{figure}

\begin{table}
  \centering\footnotesize 
  \caption{Math-to-code translations}
  \label{tab:math_to_code}
  \begin{tabular}{|l|l|}
    \hline
    \bf Code per \cref{fig:mttkrp_alg} & \bf Math per \cref{eq:mttkrp}\\ \hline
    \tt X.value(i) & $x_i$ \\ 
    \tt X.subscript(i,n) & $\ell_{in}$ \\
    \tt M.weights(j) & $\lambda_j$ \\
    \tt M[m].entry(X.subscripts(i,m),j) & $a^{(m)}(\ell_{im},j)$ \\
    \tt V.entry(k,j) & $v(k,j)$ \\
    \hline
  \end{tabular}
\end{table}

\begin{table}
  \centering\footnotesize
  \renewcommand{\arraystretch}{1.5}
  \caption{Choices for architecture-specific parameters in \cref{fig:mttkrp_alg}}
  \label{tab:arch_params}
  \begin{tabular}{|l|c|c|} \hline
    \bf Parameter & \bf CPU & \bf GPU \\ \hline
    \tt FBS & \multicolumn{2}{c|}{$\min \set{2^{\lceil\log_2(R)\rceil},32}$ } \\ \hline
    \tt vector\_size & 1 & $\min \set{\mathtt{FBS},16}$\\ \hline
    \tt team\_size & 1 & 128 / \texttt{vector\_size}\\ \hline
  \end{tabular}
\end{table}

The factor matrices and subscripts are stored in row-major order, regardless of the architecture.
The highest-level parallelism is indicated in the pseudo-code by the \inline{parallel_league_for} loop.
Each team within the league processes a mega-block of nonzeros of total size \inline{NZPT}.
The medium-level parallelism is indicated by the \inline{parallel_team_for} loop.
Each team member processes a nonzero block of size \inline{NZPTM}, which we set to 128 for all experiments.
The fine-level parallelism is indicated by the \inline{parallel_vector_for} loop.  
The vector parallelism is over a range of factor matrix columns, as each nonzero is processed.

The vector-level of parallelism over column indices exploits data locality in the factor matrices and subscripts.
However, this requires allocation of a temporary buffer \inline{tmp}.  
For efficiency, this buffer needs to be allocated in a fast memory space (such as CPU/Phi cache or GPU shared memory), which is limited in size.
Hence, we work in blocks of factor matrix columns since we assume it is not possible to store an entire factor matrix row at one time when $R$ is large.
The factor matrix column loop is blocked by a runtime determined tile size, \inline{FBS}, where each \inline{parallel_vector_for} only processes \inline{FBS} columns at one time.  Except when $R$ is small, $\mbox{\inline{FBS}}=32$ (see \cref{tab:arch_params}) ensuring fully coalesced factor matrix accesses on GPUs and ample opportunity for vectorization on CPU/Phi.
Although the tensor coordinates are reread at each iteration of the factor matrix block loop,  the tensor nonzero block size (\inline{NZPTM}) is small enough that these values fit in cache across factor matrix block iterations.
Because multiple nonzeros across multiple teams may contribute to the same entry of $\mathbf{V}$, we use \inline{atomic_add} to resolve race conditions in writes to $\mathbf{V}$.

\afterpage{\FloatBarrier}

\subsection{Implementation Using Kokkos}

We use Kokkos to implement \cref{fig:mttkrp_alg} in manner that achieves high performance and portability.  We refer the reader to the GenTen source code\footnote{\url{https://gitlab.com/tensors/genten}} to see the full implementation.

Kokkos provides a data structure called \inline{View} for storing multidimensional arrays of data with template parameters specifying the type of data, its number of dimensions, the memory space in which data is allocated, and its layout in memory.    
Using \texttt{View}, we store the indices of the sparse tensor as a two-dimensional array of ordinals and the nonzero values as a one-dimensional array of floating-point data.  The ordinal and floating-point types can be chosen while configuring GenTen, and we use \texttt{size\_t} and \texttt{double} respectively in all numerical results below.
The total size of a tensor $\X$ in memory is then $(ds_o + s_f)\cdot\mbox{nnz}(\X)$ bytes where $s_o$ and $s_f$ are the sizes in bytes of the ordinal and floating-point types, respectively, and $\mbox{nnz}(\X)$ is the number of nonzeros in $\X$.  
We also use \inline{View} to store each factor matrix as a two-dimensional array, and we store both tensor coordinates and factor matrices using Kokkos' rowwise memory layout to ensure packed/coalesced accesses of this data in the row-oriented MTTKRP algorithm discussed above.

Kokkos provides functions for the \inline{parallel_for} commands in \cref{fig:mttkrp_alg} that are specialized for each architecture according to their corresponding parallel programming libraries (e.g., OpenMP or CUDA).  In particular, the function corresponding to the \inline{parallel_vector_for} command on a CPU/Phi maps to standard \inline{for} loop that is intended to be autovectorized by the compiler.  Conversely on a GPU, the loop iterations are mapped to individual threads within a warp according to their thread index.  They each take their code bodies in the form of lambda-expressions, and accept a policy argument that describes the parallel iteration space.  Kokkos provides mechanisms for allocating and managing the per-team temporary buffer \inline{tmp}, which will be allocated in shared memory on a GPU and a fast cache on CPU/Phi.
For \inline{atomic_add}. Kokkos  calls architecture-specific built-in atomic instructions when possible.\footnote{The availability of those instructions depends upon both the hardware and the floating-point data type. For example, NVIDIA K80 GPUs have hardware atomic instructions for single-precision data but not double-precision. If hardware instructions are not available, Kokkos uses a general implementation through atomic compare-and-swap (CAS), which can be dramatically slower if there is high contention among threads.}

\section{MTTKRP with compile-time polymorphic arrays}
\label{sec:comp-time-polym}

Using Kokkos for the implementation of \cref{fig:mttkrp_alg} is portable and provides good performance on most architectures.
It makes effective use of fine-grained vector parallelism, particularly when $R$ is a multiple of the vector width of the architecture, and provides high memory-bandwidth efficiency.  Even so, it does have drawbacks compared to optimized implementations for GPU and CPU architectures since it requires allocating the temporary buffer \inline{tmp}.  On a GPU, the buffer is stored in shared memory, which is very fast compared to global memory.  However the amount of shared memory available is quite small (typically 48-96 KB and is shared by all active threads on the GPU processor) requiring the factor matrix tile size (see \cref{tab:arch_params}) to be small so that enough thread blocks can be inflight to better hide the latency of global memory accesses.
We would instead prefer to store the temporary values in registers, which are even faster to access and provide even more storage space (modern GPUs typically provide 256 KB for registers).
Such a modification would allow the tile size to be larger and therefore improve performance.
Similarly, on a CPU/Phi architecture, it would be more convenient to allocate the buffer as a simpler thread-private stack array.

\newsavebox{\firstlisting}
\begin{lrbox}{\firstlisting}
\begin{lstlisting}
template <typename Space, unsigned Length, unsigned Size, unsigned VectorSize,
          typename Enabled = void>
struct TinyVec {
  integral_nonzero_constant<unsigned,Size> sz;
  alignas(64) double v[ Length ];

  TinyVec(const unsigned size, const double x) : sz(size) {
    for (unsigned i=0; i<sz.value; ++i) v[i] = x;
  }
  
  void store_plus(double* x) const {
    for (unsigned i=0; i<sz.value; ++i) x[i] += v[i];
  }

  TinyVec& operator+=(const TinyVec& x) {
    for (unsigned i=0; i<sz.value; ++i) v[i] += x.v[i];
    return *this;
  }
    
  void atomic_store_plus(volatile double* x) const;
  TinyVec& operator=(const double x);
  TinyVec& operator*=(const double x);
  TinyVec& operator*=(const double* x);
};
\end{lstlisting}
\end{lrbox}

\newsavebox{\secondlisting}
\begin{lrbox}{\secondlisting}
\begin{lstlisting}
template <unsigned Length, unsigned Size, unsigned VectorSize>
struct TinyVec< Kokkos::Cuda,Length,Size,VectorSize,
                typename std::enable_if<Length/VectorSize == 4>::type > {
  integral_nonzero_constant<unsigned_type,Size/VectorSize> sz;
  double v0, v1, v2, v3;

  __device__ inline TinyVec(const unsigned size, const double x)
    : sz( (size+VectorSize-1-threadIdx.x) / VectorSize ) {
    v0 = v1 = v2 = v3 = x;
  }

  __device__ inline void store_plus(double* x) const {
    if (sz.value > 0) x[               threadIdx.x] += v0;
    if (sz.value > 1) x[  VectorSize + threadIdx.x] += v1;
    if (sz.value > 2) x[2*VectorSize + threadIdx.x] += v2;
    if (sz.value > 3) x[3*VectorSize + threadIdx.x] += v3;
  }

  __device__ inline TinyVec& operator+=(const TinyVec& x) {
    v0 += x.v0; v1 += x.v1; v2 += x.v2; v3 += x.v3;
    return *this;
  }

  __device__ inline void atomic_store_plus(volatile double* x) const;
  __device__ inline TinyVec& operator=(const double x);
  __device__ inline TinyVec& operator*=(const double x);
  __device__ inline TinyVec& operator*=(const double* x);

};
\end{lstlisting}      
\end{lrbox}

\begin{figure}
  \subfloat[Non-GPU architectures.]{\label{fig:tinyvec_cpu}\usebox{\firstlisting}}\\
  \subfloat[GPU/CUDA architecture when \inline{Length/VectorSize == 4}.]{\label{fig:tinyvec_cuda}\usebox{\secondlisting}}
      \caption{Example implementation of thread-local compile-time polymorphic array wrapper class (\inline{TinyVec})  for two architectures.
      The implementations of several functions, which are similar to the implementations shown, are suppressed for brevity.  The portion of the array used within the class is stored within \inline{integral_nonzero_constant<unsigned,Size>::value} which will be \inline{Size} when \inline{Size != 0}, and its constructor argument otherwise.}
      \label{fig:tinyvec}
\end{figure}

\afterpage{\FloatBarrier}

To address these shortcomings, we propose a better-performing and arguably simpler approach to portability, which we refer to as compile-time polymorphic arrays.
We extend Kokkos by introducing the class \inline{TinyVec} that has different implementations depending on the architecture, as shown in 
\cref{fig:tinyvec} %
which mirror the optimizations described above.
The current scratch-pad capability of Kokkos creates a temporary buffer that is visible to all threads within a team;
in contrast, we store the data in a thread-private manner.
The wrapper class has several template parameters, including the execution space (allowing unique implementation for each architecture through partial template specialization), the length of the array (which therefore must be a compile-time constant), a flag indicating whether all or just a portion of the array will be used, and the chosen vector size (which also must be a compile time constant).  
This class avoids the use of lambda expressions at the vector level, making it easier for the compiler to optimize and vectorize those loops.

On a CPU/Phi architecture (see \cref{fig:tinyvec_cpu}), the array will be allocated as a thread-private stack array of the given length (the factor matrix block size \inline{FBS}).  For all but the last iteration of the factor matrix block loop, the size flag indicates the \emph{compile-time} full array length will be used.  For the last iteration, the flag changes to indicate a \emph{runtime-determined} size will be used to handle the remainder term when $R$ is not evenly divisible by \inline{FBS}.  

Conversely on a GPU architecture (see \cref{fig:tinyvec_cuda}), the array data is divided up among all of the threads in a warp (determined by the vector size).  For example, if the array length is 128 and the vector size is 32, each thread holds four entries of the array.  This data could be stored in a stack array as well, which will be converted to registers by the compiler when the full array length is used.  When a dynamically sized portion of the array is used, the compiler instead stores this data in ``local'' memory which has the same high latency for accesses as global memory.  To remedy this, we directly store the data in registers by providing \emph{partial specializations} of the class based on the number of array elements per thread, and currently specializations are provided for 1-4 elements per thread.
The specialization uses typical class template specialization techniques in conjunction with Substitution Failure Is Not An Error (SFINAE)~\cite{sfinae} using \inline{std::enable_if} for the \inline{Enabled} template parameter to make the specialization available only when \inline{Length/VectorSize == 4}.
We refer the reader to the source code for complete details.

In all cases, the array wrapper provides overloaded operators for arithmetic on the array entries, where the implementation of each operator maps the operation across the entries of the array.  On a CPU/Phi architecture, these can be well-optimized and vectorized by the compiler since they are simple, lowest-level loops, and when the full array is used, have a compile-time known trip count.
On a GPU architecture, the operation is merely applied to each register variable.  Thus, each vector-parallel loop over factor matrix entries that would have been implemented through a lambda-expression is replaced by a single overloaded operator call implemented by the array wrapper.  This makes the code simpler and more understandable.  Furthermore, since shared memory is no longer used to store the temporary buffer, a larger factor matrix tile size can be used, increasing performance.  The resulting MTTKRP algorithm is displayed in \cref{fig:mttkrp_array}.

\begin{figure}
\begin{lstlisting}
parallel_league_for(league_rank=0; league_rank<league_size; ++league_rank) {

  // No shared team scratchpad!

  for (jb=0; jb<R; jb+=FBS) {
    nj = jb+FBS < R ? FBS : R-jb 
    
    parallel_team_for(team_rank=0; team_rank<team_size; ++team_rank) {
      i_offset = league_rank*NZPT + team_rank*NZPTM  
      ni = i_offset + NZPTM < P ? NZPTM : P-i_offset 
  
      for (i=i_offset; i<i_offset+ni; ++i) {
        x_val = X.value(i)     
        k = X.subscript(i,n)   
    
        // Allocate special thread-local buffer
        tmp = TinyVec(nj,x_val) // Vector op
        tmp *= &M.weights(jb) // Vector op
        
        for (m=0; m<d; ++m) 
          if (m != n) 
            tmp *= &M[m].entry(X.subscript(i,m),jb) // Vector op
  
        tmp.atomic_store_add(&V.entry(k,jb)) // Vector op

      } //i
    } // team_rank
  } // jb
} // league_rank
\end{lstlisting}
  \caption{\textbf{\VerB.} Modification of \VerA (\cref{fig:mttkrp_alg}) using \inline{TinyVec} polymorphic arrays (all code not shown is unchanged).
    In this case, we use different choices for the architecture-dependent parameters \inline{vector_size} and \inline{FBS} as shown in \cref{tab:sizes}.
    The template parameters for \inline{TinyVec} have been suppressed, including the logic for how the template parameters are determined to handle the remainder term.
    The vector operations span the range \texttt{jb} to \texttt{jn+nj-1} in the objects being accessed/written.}
\label{fig:mttkrp_array}
\end{figure}

\begin{table}
  \centering
  \caption{GPU Vector and factor block sizes for different numbers of components ($R$) for the polymorphic array MTTKRP algorithm.}
  \label{tab:sizes}
  \footnotesize
  \begin{tabular}{|l|*{13}{c|}}
    \hline
    Range for $R$ & 1 & 2 & 3 & 4 & 5--7 & 8 & 9--16 & 17--24 & 25--47 & 48 & 49--95 & 96 & 97-- \\ \hline
    \texttt{vector\_size}  & 1 & 2 & 2 & 4 & 4 & 8 & 8 & 8 & 8 & 16 & 16 & 32 & 32 \\ \hline
    \texttt{FBS} & 1 & 2 & 4 & 4 & 8 & 8 & 16 & 24 & 32 & 48 & 64 & 96 & 128 \\ \hline
  \end{tabular}
\end{table}

\afterpage{\FloatBarrier}

Conceptually our array wrapper is similar to SIMD data types studied by others~\cite{doi:10.1002/spe.1149,Karpinski:2017:HPA:3026937.3026939,Wang:2014:SPF:2568058.2568059,Kim:2017:DVC:3126908.3126941,Phipps:2017kf}, which have been used to achieve some form of outer-loop vectorization by blocking the outer loop based on the architecture's vector width and moving the vector loop to an innermost loop encapsulated by overloaded operators iterating over a statically-sized array.  Furthermore, these implementations have focused exclusively on vector CPU/Phi architectures with compile-time fixed array lengths.  The contribution here is primarily the idea of using this technique for applying fine-grained parallelism across innermost loops, for GPU and CPU/Phi architectures. This requires a polymorphic API for reading and writing data to memory as well as the ability to support run-time choice of the array length for tail loops.  We overcome performance issues peculiar to GPU architectures by storing the array entries in registers.

To achieve good performance for a wide range of $R$ values, GenTen must determine choices of the factor matrix block size and vector size.  Our logic for doing so tries to balance large factor matrix tile sizes (which reduce rereads of the tensor) with small remainders for the factor matrix loop (since the remainder portion is inherently less efficient).  The result of our logic is displayed in \cref{tab:sizes}, which shows the factor matrix block and vector sizes for ranges of $R$ values (the vector size is only relevant for the GPU architecture; it is always one on a non-GPU architecture).  On the GPU, these choices result in at most four factor matrix columns processed by each thread within a warp.  Notice that because of the greater amount of memory available for registers as opposed to shared memory, a much larger factor matrix block size is possible, compared to the scratch-pad based approach in \cref{fig:mttkrp_alg}.

%
%
%
%

\section{Permutation approach for MTTKRP}
\label{sec:perm-mttkrp}
%

%
Depending on how the nonzeros are ordered in the tensor, many threads may be trying to update partial contributions to the result factor matrix simultaneously, creating high contention for the architecture's atomic hardware.
This is particularly pronounced when there is no atomic-add instruction, such as \texttt{double} on NVIDIA Kepler architectures (e.g., K20, K40, K80), where Kokkos resorts to atomic compare-and-swap.

One approach for overcoming this challenge is to use a compressed storage format akin to the matrix compressed row storage (CRS) format, as is done in SPLATT~\cite{SmRaSiKa15}.
For a given mode $n$, the tensor is stored as a list of slices for each element of that mode.
The MTTKRP algorithm is parallelized over these slices where each thread operates on a distinct row of the resulting factor matrix,
so no atomic update is necessary.
The downside of this approach for CP-ALS is that a separate copy of the tensor is required for
each mode, since CP-ALS requires MTTKRP calculations for all modes.

In this work we introduce a new way of reducing atomic contention that is inspired by GPU implementations of the sparse matrix-vector product with sparse matrices stored in the coordinate (COO) format~\cite{Bell:2009uo}.
In this algorithm, the sparse matrix-vector product is parallelized over the matrix nonzeros.
However, instead of having each thread write its contribution using an atomic instruction, the matrix nonzeros are sorted with increasing row index.
Each thread iterates over a contiguous set of matrix nonzeros, and writes only occur when the row index changes.
When a write is necessary, all the threads within a team perform a thread-parallel segmented reduction based on the row index, combining the contributions across multiple threads without atomics.
Depending on how the algorithm is implemented, no atomic instructions may be necessary at all.

This approach cannot be directly applied to the tensor case because we want to avoid requiring $d$ copies of the tensor, each with a different sorting.
Instead we compute a permutation array for each mode that sorts the tensor nonzeros in  increasing index along that mode.
For mode $n$ MTTKRP, we iterate over the tensor nonzeros in that permuted order as opposed to the order they are stored in memory.
For a given tensor with $P$ nonzeros and $d$ modes, storing the tensor in memory requires $S = (s_r + d s_o)P$ bytes where $s_r$ and $s_i$ are the sizes of the floating-point and ordinal types, respectively.  This approach requires storing $d$ additional permutation arrays of length $P$ resulting in a total storage size of $(s_r + 2 d s_o)P < 2 S$ bytes, and therefore a little less than double the amount of memory is required to store the tensor.
The downside of this approach is the tensor nonzeros are no longer streamed from memory and are accessed in a more random fashion.
However, each tensor nonzero corresponds to $O(d\cdot F)$ floating-point operations where $F$ is the factor matrix tile size, allowing for significant reuse of those values.

This idea requires only small modifications of the MTTKRP kernel implementation so that each tensor nonzero index is extracted from the permutation array. %
The same logic is used for determining the vector and factor matrix tile sizes as shown in \cref{tab:sizes}.
Each thread within a team iterates over a given block size of tensor nonzeros and writes its contribution to the resulting factor matrix only when the mode-$n$ coordinate changes.
This must be an atomic-write if the mode-$n$ index is equal to the first or last index of the block (since another thread may be writing to the same row); otherwise, it is a regular (non-atomic) write.
Determining when and what kind of write should happen results in most of the changes to the kernel implementation, as shown in \cref{fig:mttkrp_perm}.
We could reduce contributions across threads within a team as is typically done in the sparse matrix-vector product algorithm to reduce the frequency of atomic writes even further; however this requires synchronization within each team, which appears to offset any potentially improved performance induced by the inter-team reduction.

\begin{figure}
\begin{lstlisting}
parallel_team_for(team_rank=0; team_rank<team_size; ++team_rank) {
  i_offset = league_rank*NZPT + team_rank*NZPTM  
  ni = i_offset + nzptm < P ? nzptm : P-i_offset 
  val = TinyVec(nj,0.0) // accumulation buffer for a row, initialize to zero
  row_prev, first_row = -1 // used to track when to store

  // Loop over tensor nonzeros in block, in order wrt mode n
  for (i=i_offset; i<i_offset+ni; ++i) {
    p = X.getPerm(i,n)     // index of i-th nonzero in permuted ordering 
    x_val = X.value(p)     // value for nonzero p
    row = X.subscript(p,n) // mode-n index for nonzero p

    // Is this the first row in a block? 
    if (i == i_offset) {
      first_row = row
      row_prev = row
    }
      
    // Detect change in row index
    if (row != row_prev) {           
      // Sum the result into the appropriate entries in V
      if (row_prev == first_row) // First row needs atomic operation
        val.atomic_store_plus(&V.entry(row_prev,jb)) 
      else // Row owned entirely by this process, no atomic operation
        val.store_plus(&V.entry(row_prev,jb)) 

      //Reset for next row
      val = 0            
      row_prev = row
    }
        
    tmp = TinyVec(nj, x_val)
    tmp *= &M.weights(jb)
  
    for (unsigned m=0; m<d; ++m) 
      if (m != n)
        tmp *= &(M[m].entry(X.subscript(p,m),jb))
        
    // Accumulate sum across nonzeros until row index changes
    val += tmp

    // Last row needs atomic operation
    if (i == offset+ni-1) 
      val.atomic_store_plus(&V.entry(row,jb)) // Vector operation

  } // i
} // team_rank
\end{lstlisting}
\caption{\textbf{\VerC.} Modification of \VerB (\cref{fig:mttkrp_array}) to use permuted indexing and avoid most atomic operations  (all code not shown is unchanged).}
\label{fig:mttkrp_perm}
\end{figure}

There is a small preprocessing cost associated with this method to compute the permutation array for each tensor mode.
\genten can compute the permutation array using parallel sorting routines provided by Kokkos, as well as
Thrust\footnote{\url{http://docs.nvidia.com/cuda/thrust/index.html}} for GPU architectures and the Intel Parallel Stable Sort\footnote{\url{https://software.intel.com/en-us/articles/a-parallel-stable-sort-using-c11-for-tbb-cilk-plus-and-openmp}} for OpenMP-based architectures.

%

%
%
%
%

\afterpage{\FloatBarrier}

\section{Numerical results}
\label{sec:results}
%

We investigate the performance of the different proposed portable parallel implementations of CP-ALS.
Our results are generated with the publicly available \genten library on four different architectures (both CPU and GPU).
The specific architectures and compilers are listed in \cref{table:arch},
with the architecture- and compiler-specific optimization flags specified by Kokkos.

\subsection{Artificial data and scalability studies}
\begin{figure}
  \centering
  \subfloat[\label{cpals_time_full}%
  Time for each architecture, showing MTTKRP is the most expensive operation.
  (Using 100\% thread capacity.)
  ]%
  {\includegraphics[width=0.5\textwidth]{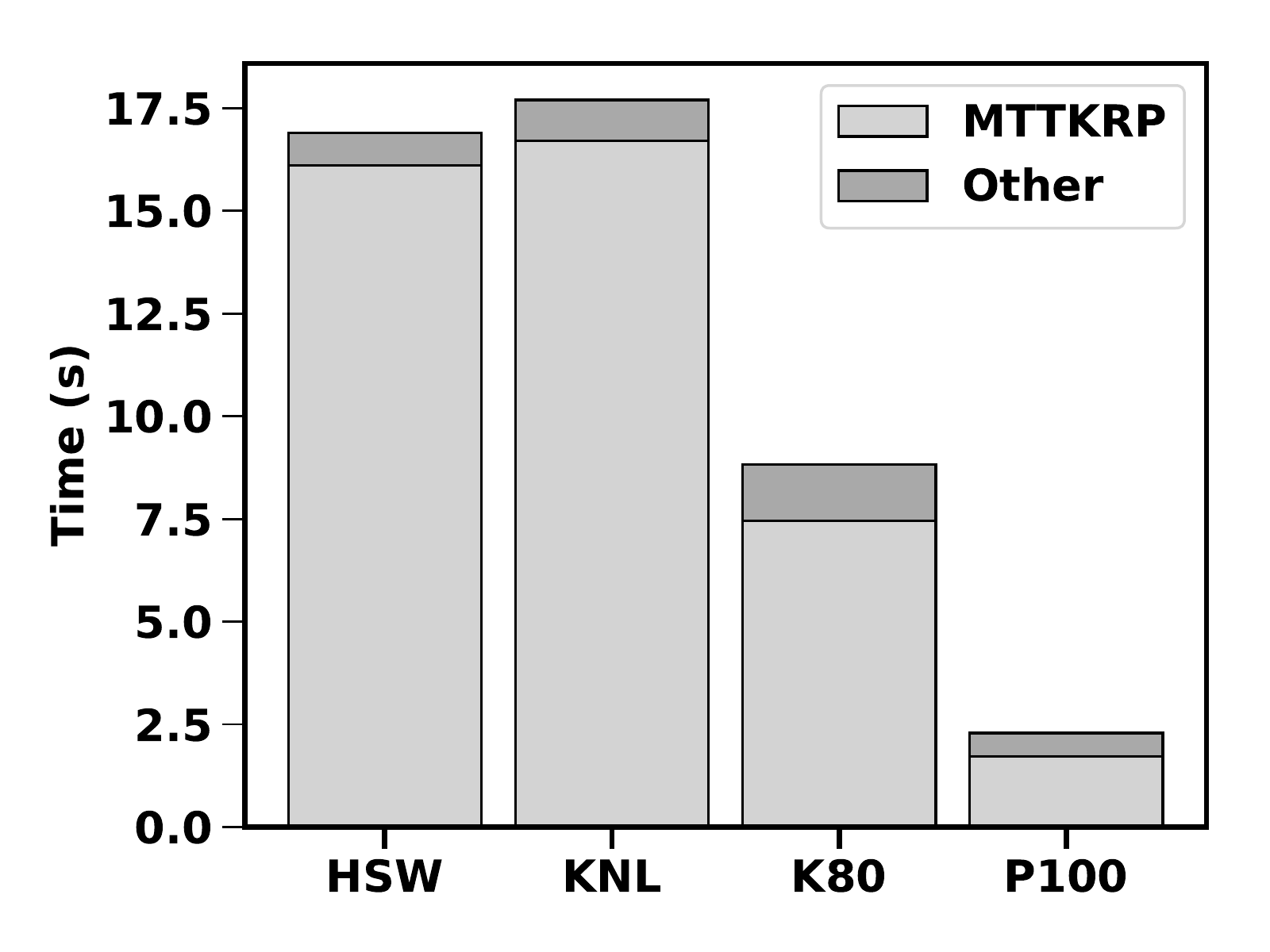}}
  \subfloat[\label{cpals_time_scale}%
  For Haswell CPU and KNL platforms, showing time as the numbers of threads increases to the  maximum available.
  (The number of threads used cannot be varied on GPU platforms.)
  ]%
  {\includegraphics[width=0.5\textwidth]{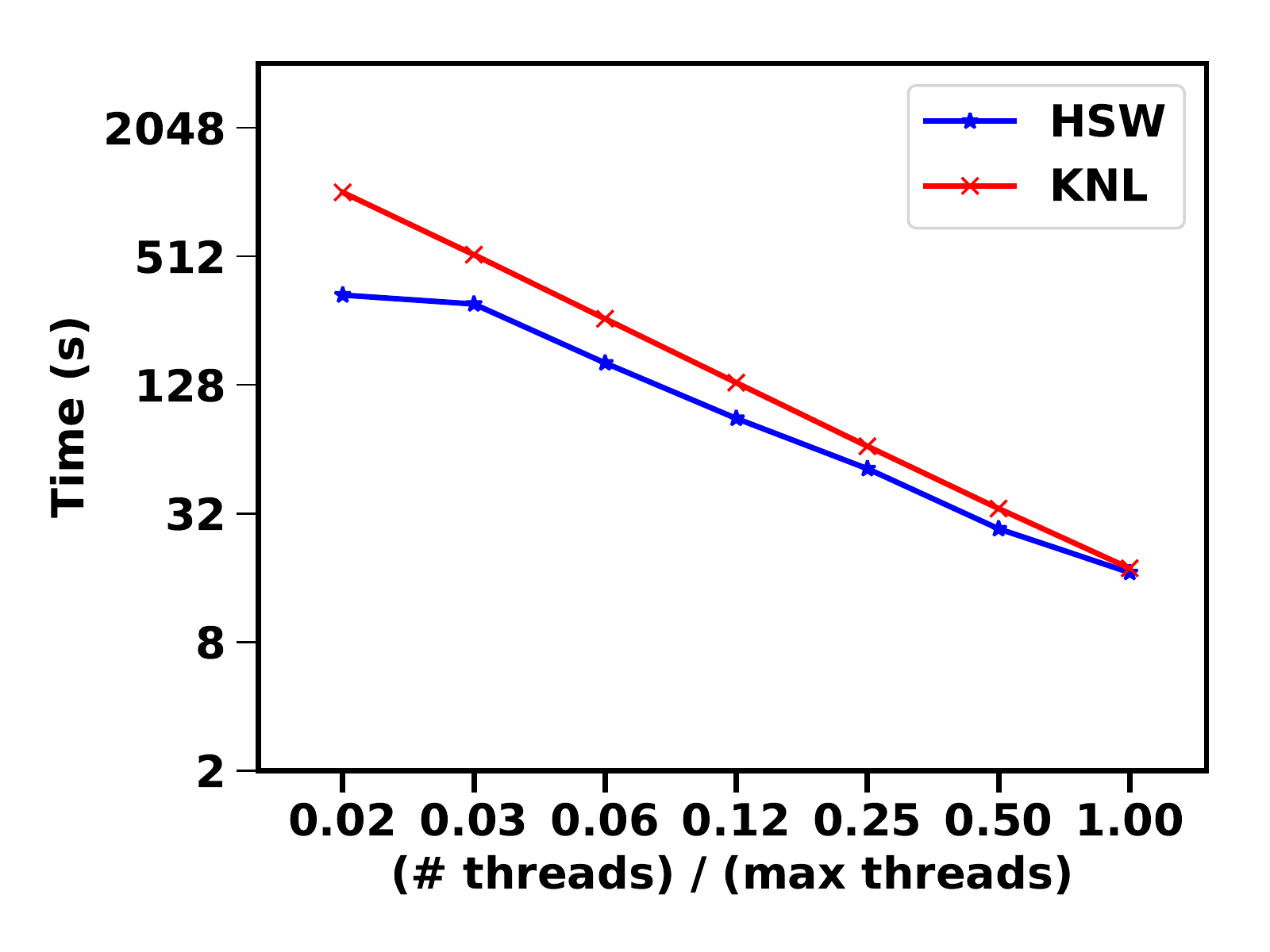}}
  \caption{Total run times on different architectures for ten iterations of CP-ALS using \VerB on a tensor of size $30\mbox{K}\times40\mbox{K}\times50\mbox{K}$ with $10\mbox{M}$ nonzeros and $R=128$.
  }
  \label{fig:cpals_time}
\end{figure}

We first consider running the full CP-ALS method on a synthetic three-dimensional tensor of size $30,000\times40,000\times50,000$ with 10 million nonzeros placed randomly throughout the tensor.
For each architecture, \Cref{fig:cpals_time} displays the total run time for 10 iterations of CP-ALS with $R=128$ factor components using \VerB (\cref{fig:mttkrp_array}).
\Cref{cpals_time_full} displays the total run-time for all four architectures at full machine capacity.
First, MTTKRP takes most of the CP-ALS computation time, regardless of architecture.
Second, the Pascal P100 GPU is substantially faster than the other three.
As compared to single Haswell core, we see about a 20-fold speedup on Haswell and KNL, a 38-fold speedup on the K80, and a 147-fold speedup on the P100.
For the Haswell and KNL architectures, \cref{cpals_time_scale}  varies the number of threads up to the maximum number used (64 and 256 respectively), and plots the total run-time in seconds against the fraction of the total number of threads used.\footnote{On both architectures these results were generated with \texttt{OMP\_PROC\_BIND=close} and \texttt{OMP\_PLACES=threads} OpenMP thread-binding environment variables set.} 
These plots demonstrate good thread scalability on the Haswell and KNL architectures.
For the GPU architectures, it is not possible to vary the number of threads.

\begin{table}[!htb]
  \renewcommand{\arraystretch}{1.25}
  \centering\footnotesize
  \caption{Computational parallel architectures used for experimental results. While the KNL architecture supports up to 272 threads, at most 256 threads (64 cores) were used for our experiments.  The KNL was placed in cache mode, where the high-bandwidth memory (HBM) is used as a last-level cache for main memory.}
  \label{table:arch}
  \begin{tabulary}{\textwidth}{llLll}
    \toprule
     &  & Architecture & Thread  & \\[-1ex]
    Type & Name & Description & Parallelism & Compiler\\
    \midrule
    CPU & Haswell & Intel Xeon E5-2698v3 CPU, 2.3 GHz, 2 sockets, 16 cores/socket, 2 threads/core,  max~32~threads & OpenMP & Intel 17.0 \\ 
    CPU & KNL  & Intel Xeon Phi 7250, 68 cores, 4~threads/core, HBM in cache mode, use~max~256~threads & OpenMP & Intel 18.2 \\
    GPU & K80  & NVIDIA Kepler K80 GPU & CUDA & NVCC 9.2 \\
    GPU & P100 & NVIDIA Pascal P100 GPU & CUDA & NVCC 9.2 \\
    \bottomrule
  \end{tabulary}
\end{table}

\begin{figure}
  \centering
  \subfloat[\label{mttkrp_both_full}
  Time for each architecture, showing \VerC is faster in every case. (Using 100\% thread capacity.)
  ]%
  {\includegraphics[width=0.5\textwidth]{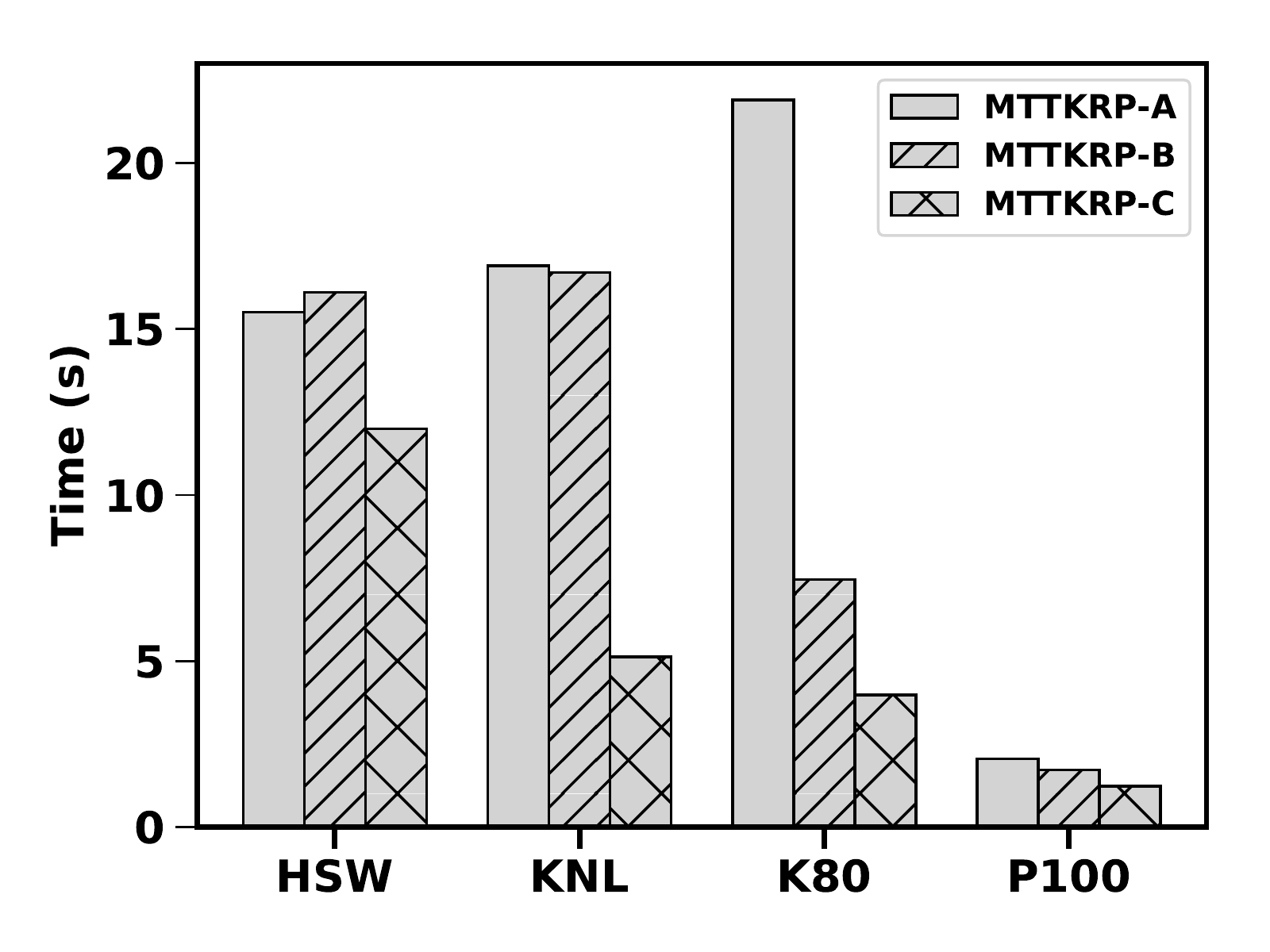}}
  \subfloat[\label{mttkrp_both_scale}
  For Haswell CPU and KNL platforms, showing time as the numbers of threads increases to the  maximum available.
  (The number of threads used cannot be varied on GPU platforms.)
  ]%
  {\includegraphics[width=0.5\textwidth]{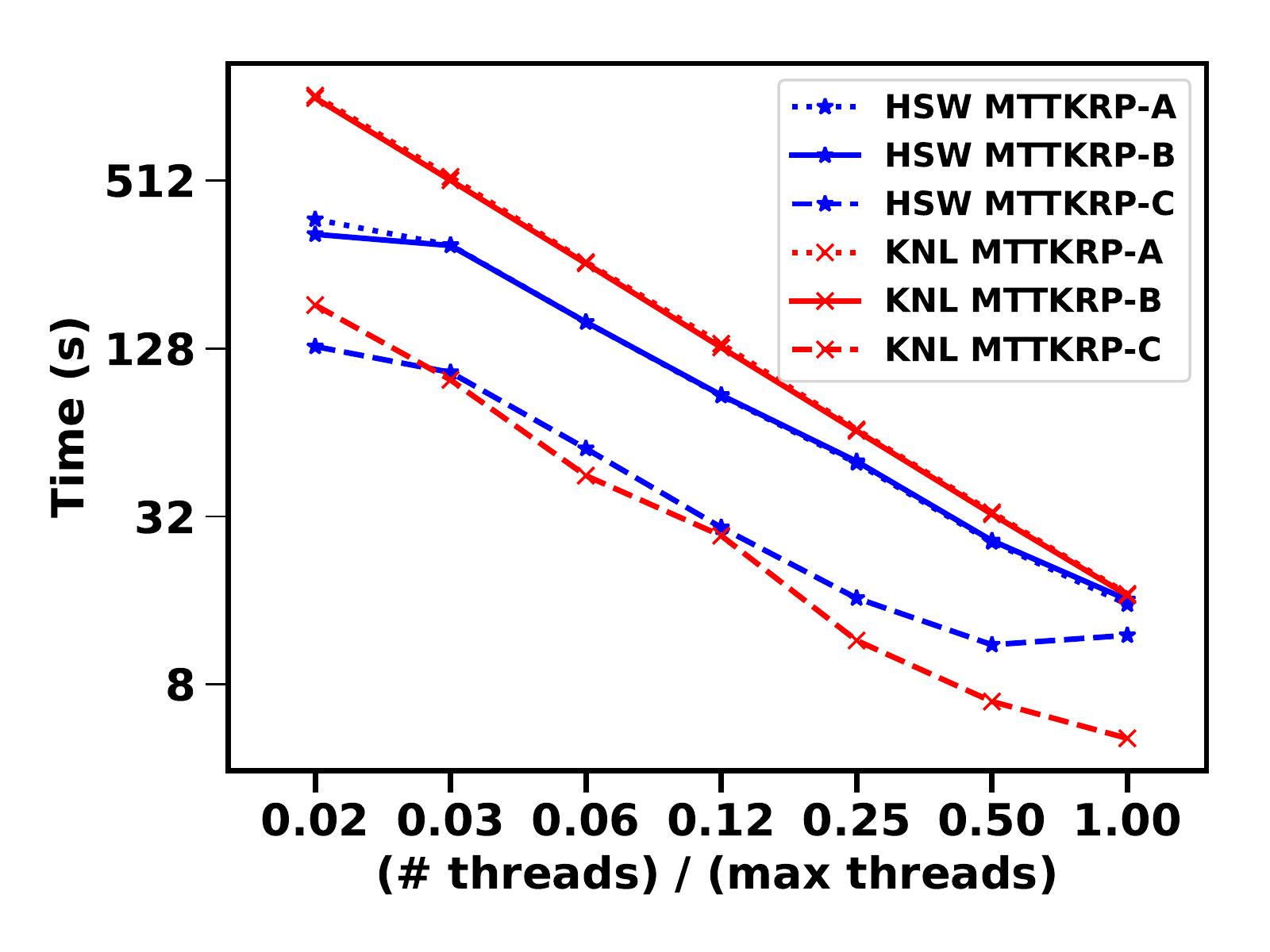}}
  \caption{Comparison of total run times for all three MTTKRP algorithms, on different architectures, for for ten iterations of CP-ALS on
    a tensor of size $30\mbox{K}\times40\mbox{K}\times50\mbox{K}$ with $10\mbox{M}$ nonzeros and $R=128$ .
  }
  \label{fig:mttkrp_both}
\end{figure}

In \cref{fig:mttkrp_both}, on the same data tensor, we compare \VerA (\cref{fig:mttkrp_alg}), \VerB (\cref{fig:mttkrp_array}) and \VerC (\cref{fig:mttkrp_perm}) in terms of total MTTKRP time over 10 iterations of CP-ALS.   We see that the use of the polymorphic arrays (\VerB) has little effect on performance for Haswell and KNL in comparison to the base \VerA, but substantially improves performance on the K80 GPU.  Furthermore, we see the permutation-based approach (\VerC) is faster on all architectures, substantially so on the Haswell, KNL and K80 architectures, but with not much difference on the P100.
Clearly, atomic-write throughput is a bottleneck in MTTKRP performance, so \VerC is an improvement.
We see little difference for the P100 because it has very fast double-precision atomic throughput.
These timings do not include the additional setup cost associated with the permuted algorithm to compute the $d$ permutation arrays, which is investigated later in \cref{table:sorting_cost}.

\begin{figure}[h]
  \centering
  \subfloat[Measured peak bandwidth for each architecture, as determined by STREAM.\label{table:mttkrp_bandwidth}]{
    \begin{tabular}{cc}
      \midrule
      Architecture & Bandwidth (GB/s) \\
      \midrule
      HSW & 96  \\
      KNL & 294  \\
      K80 & 177  \\
      P100 & 544  \\
      \midrule
    \end{tabular}
  }\\
	\subfloat[\label{fig:mttkrp_bandwidth_hsw} HSW]{\includegraphics[width=0.5\textwidth]{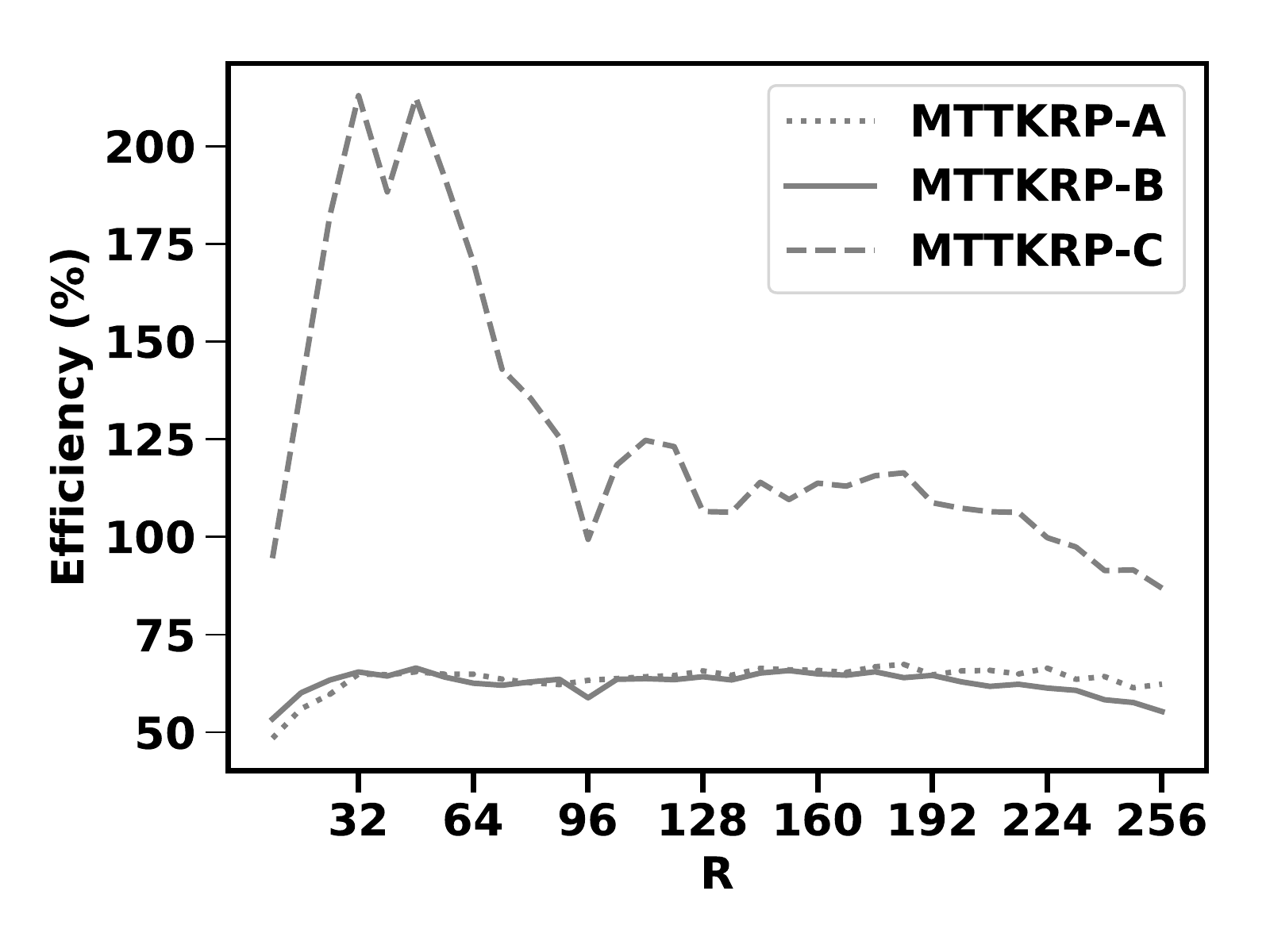}}
	\subfloat[\label{fig:mttkrp_bandwidth_knl} KNL]{\includegraphics[width=0.5\textwidth]{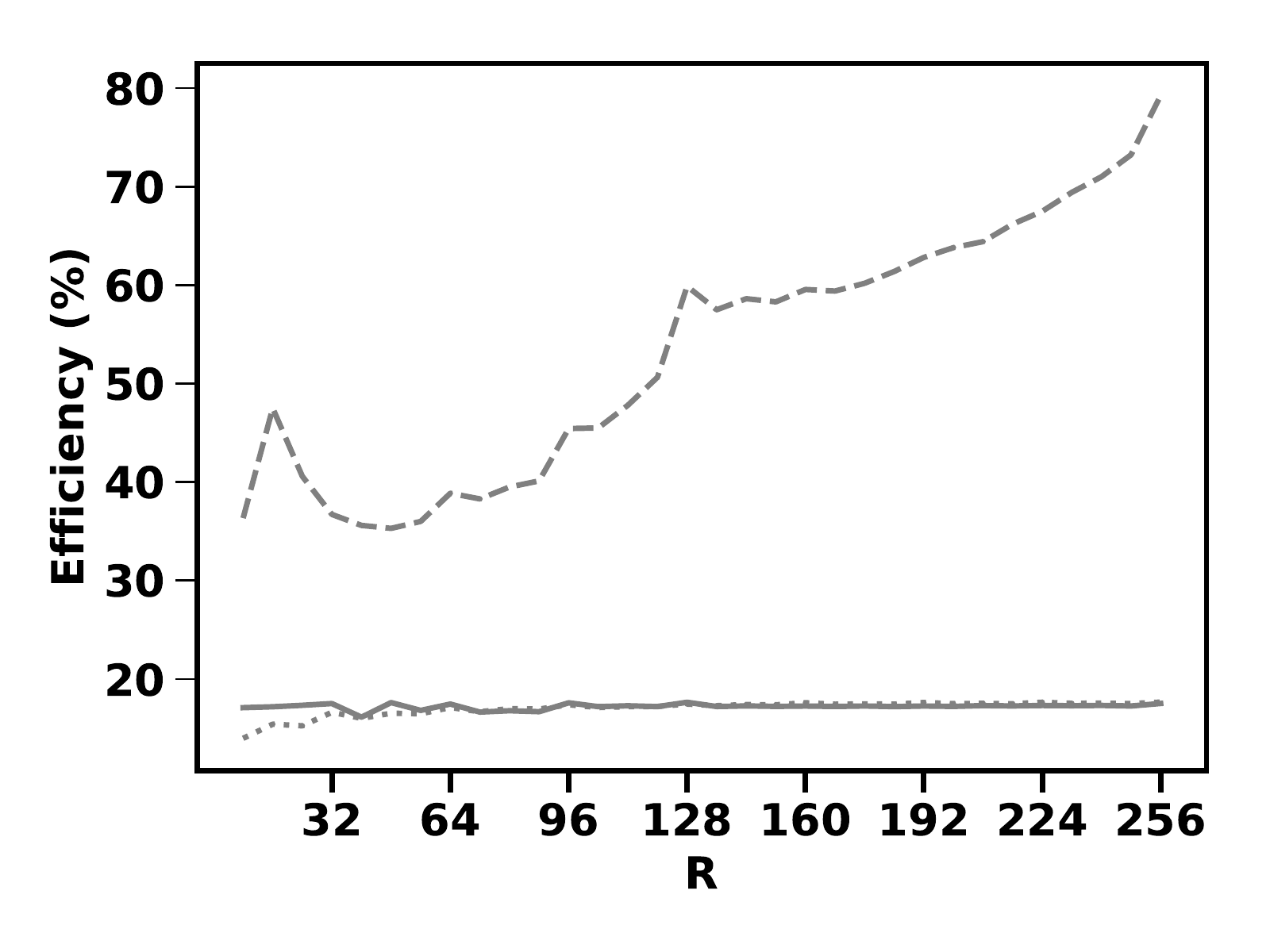}} \\
	\subfloat[\label{fig:mttkrp_bandwidth_k80} K80 GPU]{\includegraphics[width=0.5\textwidth]{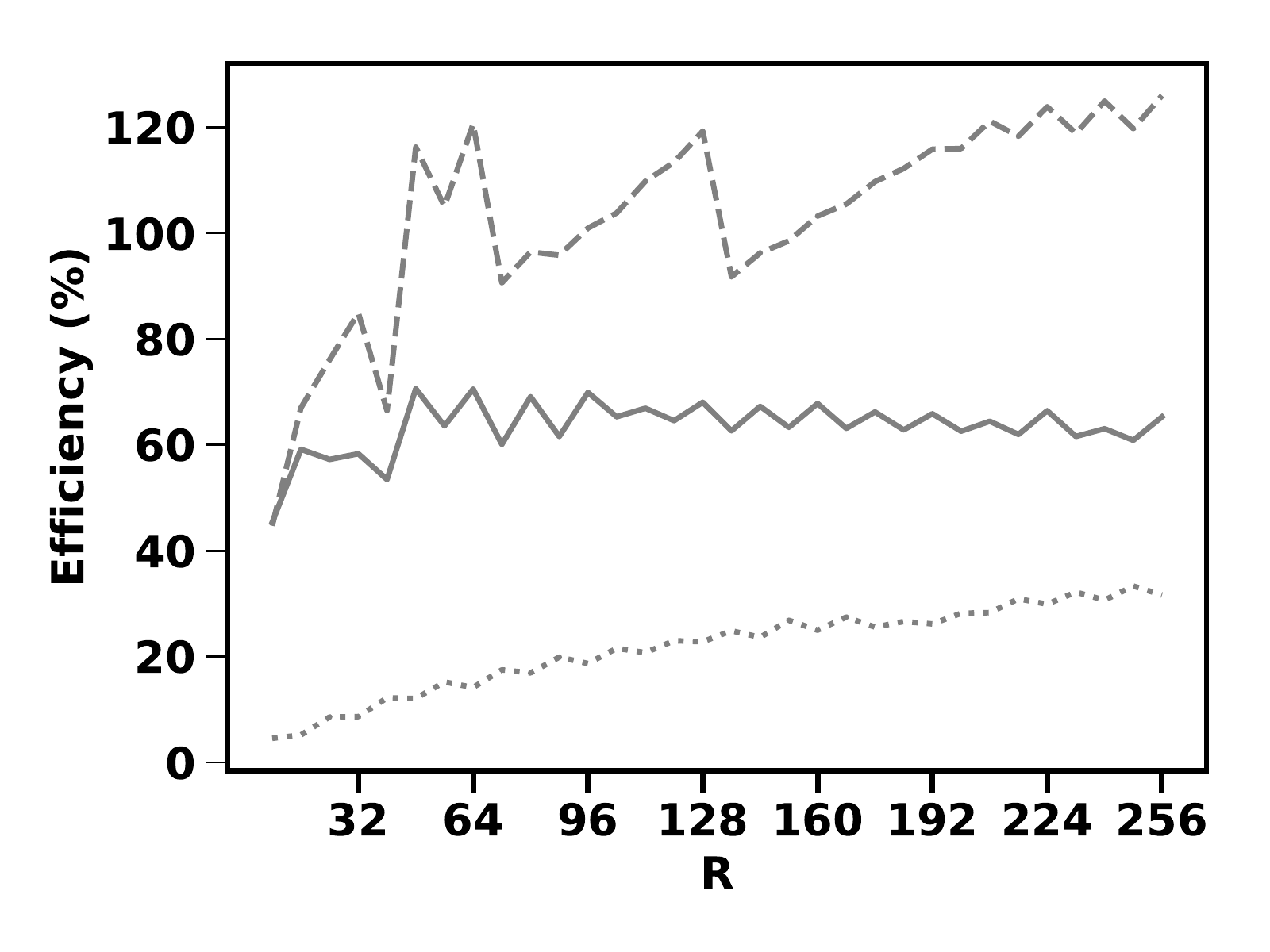}}
	\subfloat[\label{fig:mttkrp_bandwidth_p100} P100 GPU]{\includegraphics[width=0.5\textwidth]{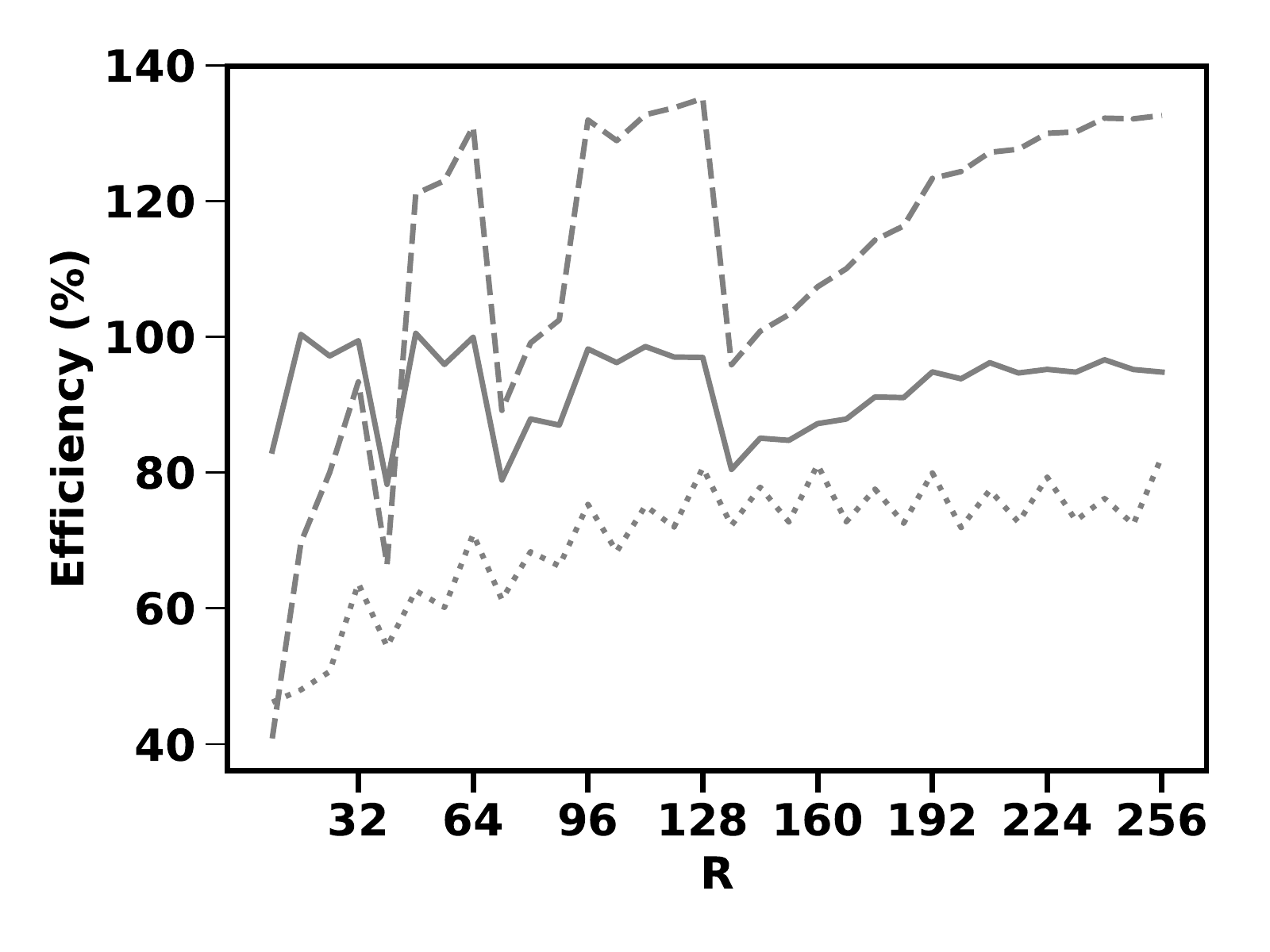}}
	\caption{MTTKRP memory bandwidth as a percentage of peak bandwidth. %
            The tensor is of size $30\mbox{K}\times40\mbox{K}\times50\mbox{K}$ with $10\mbox{M}$ nonzeros
            and we vary the number of factor components $R$ (x-axis).
            The computation used 64-bit floating point and ordinal values.
        	\label{fig:mttkrp_bandwidth}}
\end{figure}

To understand how our performance compares to what we might expect, we do some simple analysis.
If we assume all tensor and factor components values are read from main memory with no caching and no thread contention for atomic-writes, then MTTKRP is a bandwidth-limited computation driven by the peak bandwidth of each architecture.  To determine the peak bandwidth, we used the STREAM benchmark~\cite{McCalpin1995} with the results shown in \cref{table:mttkrp_bandwidth}.  We estimate the bandwidth for MTTKRP as
\begin{equation}
(d R+3) s_r + d s_o)P/t
\end{equation}
where $s_r$ is the size (in bytes) of the floating-point type, $s_o$ is the size of the ordinal type, $P$ is the number of tensor nonzeros, and $t$ is the total time of the MTTKRP operation.
The measured bandwidth for all three versions of MTTKRP on all four architectures for $R\in[8,256]$ with a stride of 8 is shown in \cref{fig:mttkrp_bandwidth} as a percentage of the peak bandwidth in \cref{table:mttkrp_bandwidth}.  For all four architectures, the bandwidth efficiency of \VerC is better since the cost of the atomic writes is substantially reduced, making the estimate a better model of the actual MTTKRP calculation.
The measured bandwidth can be greater than 100\% because of cached reads of the factor matrix entries (even on the GPU architectures, some of the data reads are automatically cached in the L2 and constant caches).
For KNL, the bandwidth efficiency only approaches 100\% for very large $R$ and is indicative of the general difficulty of achieving full memory bandwidth on this architecture.
For the GPU architectures, we also see the efficiency can be quite sensitive to $R$, providing the best performance when $R$ is a multiple of 32, which is due to the methodology for computing the vector and factor matrix tile sizes in \cref{tab:sizes}. Finally we see the use of \inline{TinyVec} in \VerB substantially improves performance as compared to \VerA for the GPU architectures by allowing for larger factor matrix block sizes.  No improvement is seen on HSW and KNL, primarily because the original algorithm is clearly limited by atomic contention rather than memory throughput.

\subsection{Real-world data and comparisons to SPLATT}

We study the MTTKRP performance on several tensors available from the Formidable Repository of Open Sparse Tensor and Tools (FROSTT)~\cite{frosttdataset}.
We selected tensors to be as large as possible, so long as they still fit within the limited memory available on the K80 GPU (12 GB).
A summary of the tensors and their sizes is given in  \cref{table:tensors}.

We compare two versions of our method (\VerB and \VerC) to the state of the art, SPLATT's CSF-based algorithms~\cite{Smith:2015ff}.
We use two versions of SPLATT:  mutexes (SPLATT-M) and tiling without mutexes (SPLATT-T), both with the default of two CSF modes.
SPLATT does not have a GPU implementation, so we only include it  
for the two OpenMP-based architectures (Haswell and KNL).
\Cref{table:mttkrp_time} shows the results based on the total MTTKRP time for 10 CP-ALS iterations with $R=16$ factor components.
The timing results shown here are aggregated over all modes of the tensor, even though significant variation in performance is often observed for different modes of a given tensor.
While performance of the COO-based MTTKRP is relatively insensitive to the length of each mode (since it parallelizes over nonzeros and not factor matrix rows), it is quite sensitive to the ordering of nonzero coordinates in each mode (since this affects the amount of atomic contention).  However the permutation-based approach by design attempts to mitigate this.

\begin{table}
  \centering\footnotesize
  \caption{Results on real-world tensor data sets.}
  \subfloat[Real-world tensor data sets from FROSTT.\label{table:tensors}]{
\begin{tabular}{cccccc}
\midrule
Name & Order & Dimensions & Nonzeros \\
\midrule
LBNL & 5 & $1.6\mbox{K} \times 4.2\mbox{K} \times 1.6\mbox{K} \times 4.2\mbox{K} \times 868\mbox{K}$ & 1.7M \\
Uber & 4 & $183 \times 24 \times 1.1\mbox{K} \times 1.7\mbox{K}$ & 13M \\
Enron & 4 & $6.0\mbox{K} \times 5.7\mbox{K} \times 244\mbox{K} \times 1.2\mbox{K}$ &  54M \\
VAST & 5 & $165\mbox{K} \times 11\mbox{K} \times 2 \times 100 \times 89$ & 26M \\
NELL2 & 3 & $12\mbox{K} \times 9.1\mbox{K} \times 29\mbox{K}$ & 77M \\
Delicious & 4 & $532\mbox{K} \times 17\mbox{M} \times 2.5\mbox{M} \times 1.4\mbox{K}$ & 140M \\
\midrule
\end{tabular}}
\\
\subfloat[Total MTTKRP time (in seconds) for ten iterations with $R{=}16$ for each data tensor, architecture, and method.  Due to the increased storage requirements of the permuted approach, global memory was exhausted for \VerB for the Delicious tensor on the K80 and P100 architectures.
\label{table:mttkrp_time}]{
\begin{tabular}{cccccccc}
\midrule
Arch. & Method & LBNL & Uber & Enron & VAST & NELL2 & Delicious \\
\midrule
HSW & \VerB & {16.6} & {8.3} & {45.7} & {284.0} & {42.8} & {57.4} \\
 & \VerC & \bf {0.4} & {0.3} & {5.7} & {6.3} & {7.0} & \bf {22.7} \\
 & SPLATT-M & {0.8} & \bf {0.2} & \bf {1.1} & {26.1} & \bf {1.3} & {23.4} \\
 & SPLATT-T & {0.5} & {1.6} & {10.4} & \bf {3.8} & {1.4} & {71.2} \\
\\
KNL & \VerB & {41.2} & {9.2} & {108.0} & {1910.0} & {82.4} & {79.1} \\
 & \VerC & \bf {0.4} & \bf {0.2} & {2.8} & {16.1} & \bf {3.2} & \bf {12.9} \\
 & SPLATT-M & {2.0} & {1.2} & \bf {2.6} & {150.5} & {9.0} & {47.7} \\
 & SPLATT-T & {1.0} & {5.1} & {41.3} & \bf {6.5} & {15.6} & {219.1} \\
\\
K80 & \VerB & {72.4} & {5.6} & {218.0} & {2010.0} & {55.2} & {30.3} \\
 & \VerC & \bf {0.5} & \bf {0.5} & \bf {9.4} & \bf {7.4} & \bf {11.6} &  --  \\
\\
P100 & \VerB & {0.4} & {0.2} & {2.0} & {4.1} & \bf {1.6} & {4.0} \\
 & \VerC & \bf {0.1} & \bf {0.1} & \bf {1.7} & \bf {1.5} & {1.8} &  --  \\
\midrule
\end{tabular}
}
\\
\subfloat[Sorting cost for the permutation-based MTTKRP approach scaled by the average (permuted) CP-ALS iteration time with $R {=} 16$.
\label{table:sorting_cost}]{
\begin{tabular}{ccccccc}
\midrule
Architecture & LBNL & Uber & Enron & VAST & NELL2 & Delicious \\
\midrule
HSW & 0.6 & 4.3 & 9.2 & 5.3 & 6.2 & 4.0 \\
KNL & 1.0 & 5.5 & 8.6 & 1.3 & 7.6 & 4.9 \\
K80 & 1.1 & 3.0 & 3.3 & 3.3 & 2.6 &  --  \\
P100 & 0.5 & 1.3 & 4.2 & 3.6 & 4.6 &  --  \\
\midrule
\end{tabular}
}
\end{table}

If we compare just \VerB and \VerC, then \VerC (the permuted approach) is up to two orders of magnitude faster.
The greater differences for real-world data tensors is due to higher atomic contention.
This is most severe on the K80 architecture due to its lack of double-precision atomic instructions;
conversely, it is least severe on the P100 architecture because of its support for double-precision atomic instructions.
Unfortunately, the cost of storing the permutations exhausted the global memory on the K80 and P100 for the largest tensor (with 140M nonzeros).

If we compare to SPLATT on HSW, our permuted approach is faster in two cases and never more then six times slower.
On KNL, our approach is more than three times faster for the largest tensor, and never worse than three times slower.
Overall, we claim that the performance of our code is comparable to that of SPLATT (i.e., same order of magnitude) while
having the advantage of being portable.

The permutation has preprocessing cost, just as SPLATT has a preprocessing cost for its CSF data structures.
We show the cost of the sorting time required for the \VerC, scaled by the average CP-ALS iteration time using \VerC, for each tensor and architecture in \cref{table:sorting_cost}.
As described above, we use Thrust for the sorting on the GPU architectures and the OpenMP-based Intel Parallel Stable Sort on Haswell and KNL.
The sorting time is less than the cost of ten CP-ALS iterations, which is relatively small and usually worthwhile given the improvement on architectures that are sensitive to atomic writes (i.e., HSW, KNL, K80).

%

%
%
%
%

\section{Conclusions}
\label{sec:conclusions}
%

In this paper we describe a portable and performant implementation of MTTKRP for sparse tensors on emerging computer architectures, including multicore CPUs, manycore Intel Xeon Phis, and NVIDIA GPUs.
For a sparse tensor stored in coordinate format, 
we showed how to arrange the loops to achieve fine-grained parallelism,
in \cref{fig:mttkrp_alg}.
The portable implementation is primarily facilitated by the Kokkos library, which provides data structures and abstractions that enable performance on multiple architectures.
One of the complications of our implementation is that each thread requires
its own temporary storage. Such an allocation is limited in Kokkos, 
so we introduced \inline{TinyVec}, an extension of the Kokkos framework for polymorphic data arrays that stores the temporary data in registers
and whose length is parameterized by an architecture-dependent
compile-time constant. The resulting algorithm is shown in \cref{fig:mttkrp_array}.
To avoid atomic operations, we do some preprocessing so that we can loop through the coordinates of each mode in increasing order through the use of \emph{permutation arrays}.
This doubles the storage for the indices, but the increase in performance
for the version shown in \cref{fig:mttkrp_perm} can be considerable.
Our implementation of CP-ALS using our improvements to MTTKRP is available in the open-source software package \genten.

We studied the performance of \genten on a variety of contemporary architectures and demonstrated that \genten's MTTKRP is efficient on all of these architectures by comparing to the expected computational bandwidth.
We also compared the performance of \genten's MTTKRP to SPLATT on CPU and KNL platforms using several realistic data tensors from FROSTT, demonstrating comparable or better performance for the permutation-based MTTKRP algorithm.  

Future work with \genten will involve the incorporation of distributed memory parallelism to enable analysis of larger tensors, as well as generalization of the algorithms to be applicable to more general categories of data tensors (such as count and binary data) using the generalized CP method introduced in \cite{arXiv-HoKoDu18}.
We will also investigate approaches for performing the necessary tensor sorting operations required by the permutation-based MTTKRP algorithm while the tensor is being read from disk to eliminate this extra cost (by, e.g., having one thread read the data from disk and have one or more threads sort it).
Following some of the ideas in SPLATT \cite{Smith:2017em}, 
performance on KNL architectures can be improved
by placing the factor matrices in high-bandwidth (HBM) memory (and putting the KNL in so-called flat mode) and most other data in main memory.
(This optimization does not appear to be present in the public version of SPLATT used in our results).
Results in \cite{Smith:2017em} suggest this could improve performance by as much as a factor of two.
We are also exploring other approaches to addressing atomic contention within MTTKRP, such as the use of thread-private copies of the resulting factor matrix that are later reduced across threads (similar to the approach presented in \cite{Smith:2017em}).
Preliminary work in this direction suggests it can be beneficial to the coordinate-based MTTKRP algorithm, as long as the number of rows of the factor matrix or the number of threads is not too large.
Ultimately, we are likely to converge to a hybrid approach that  combines atomic and reduction approaches.
%
%
%
%

%
%
%
%
%
%

\section*{Acknowledgments}
We thank the anonymous referees for critical feedback on this work,
resulting in substantial improvements in the presentation.
We graciously acknowledge funding support from the Lab Directed Research and Development (LDRD) program and Sandia National Laboratories.

\bibliographystyle{siamplain}


\end{document}